\documentclass[aps,preprint,prd,showpacs,nofootinbib]{revtex4}
\usepackage{amsmath,amssymb}
\usepackage{graphicx,subfigure}
\usepackage{color,multirow}
\usepackage[colorlinks,linkcolor=magenta,anchorcolor=cyan,citecolor=blue,plainpages=false]{hyperref}

\hypersetup{colorlinks=true,
    breaklinks=true,
    pdfstartview=Fit,
    linkcolor=blue,
    citecolor=blue,
    urlcolor=blue}

\def\be{\begin{equation}}
    \def\ee{\end{equation}}
\def\ba{\begin{eqnarray}}
    \def\ea{\end{eqnarray}}

\begin{document}

\title{Trapped early dark energy }

\author{Hao Wang$^{1,2} $\footnote{\href{wanghao187@mails.ucas.ac.cn}{wanghao187@mails.ucas.ac.cn}}}
\author{Yun-Song Piao$^{1,2,3,4} $ \footnote{\href{yspiao@ucas.ac.cn}{yspiao@ucas.ac.cn}}}

    \affiliation{$^1$ School of Fundamental Physics and Mathematical
        Sciences, Hangzhou Institute for Advanced Study, UCAS, Hangzhou
        310024, China}

    \affiliation{$^2$ School of Physics Sciences, University of
        Chinese Academy of Sciences, Beijing 100049, China}

    \affiliation{$^3$ International Center for Theoretical Physics
        Asia-Pacific, Beijing/Hangzhou, China}

    \affiliation{$^4$ Institute of Theoretical Physics, Chinese
        Academy of Sciences, P.O. Box 2735, Beijing 100190, China}

    \begin{abstract}

As a prospective resolution of the Hubble tension, early dark
energy (EDE) suffers from the coincidence problem, why EDE is
active just at matter-radiation equality (equivalently why the
slope of EDE potential is required to approximately equal to the
Hubble parameter at that time). In this paper we present a
dark-matter-trapped EDE mechanism, by which the bound on the slope
of EDE potential can be relieved. We show how this mechanism can
work, and discuss the possibility that after inflation ended EDE
can settle down at the initial conditions required by
observations.

    \end{abstract}

    \maketitle

There is a 5$\sigma$ discrepancy between the Hubble constant,
$H_0=67.37\pm0.54$km/s/Mpc, inferred by the Planck collaboration
\cite{Planck:2018vyg} using cosmic microwave background (CMB)
based on $\Lambda$CDM and, $H_0=73.04\pm1.04$km/s/Mpc, measured
locally by the SH0ES team \cite{Riess:2021jrx}. This well-known
Hubble tension, indicating new physics beyond $\Lambda$CDM, has
gathered wide attention over the past several years
\cite{Verde:2019ivm,Knox:2019rjx,DiValentino:2020vnx,DiValentino:2021izs,Perivolaropoulos:2021jda}.


As a prospective resolution to the Hubble tension, early dark
energy (EDE) has been studied intensively
\cite{Karwal:2016vyq,Poulin:2018cxd,Smith:2019ihp,Agrawal:2019lmo,Alexander:2019rsc,Ye:2020btb,Ye:2020oix,Lin:2019qug,Sakstein:2019fmf,Niedermann:2019olb,Braglia:2020iik,Braglia:2020bym,Braglia:2020auw,Kaloper:2019lpl},
see \cite{Poulin:2023lkg,Vagnozzi:2023nrq} for recent review and
Ref.\cite{McDonough:2023qcu} for current observational
constraints. In corresponding models, EDE is non-negligible only
shortly before recombination, which thus suppresses the sound
horizon and lifts $H_0$ to a higher value than that in
$\Lambda$CDM. Particularly, an anti-de Sitter (AdS) phase existing
around recombination can efficiently enhance the EDE contribution
and lead a bestfit $H_0\sim 73$km/s/Mpc
\cite{Ye:2020btb,Ye:2020oix}.
There have been lots tests to EDE based on different datasets,
e.g.
\cite{Chudaykin:2020acu,Chudaykin:2020igl,Jiang:2021bab,Hill:2021yec,LaPosta:2021pgm,Ye:2022afu,Simon:2022adh,Cruz:2022oqk,Ye:2023zel,Cruz:2023lmn,Gsponer:2023wpm},
see also recent \cite{Efstathiou:2023fbn}, However, it should be
mentioned that the rise of $H_0$ is usually accompanied with the
exacerbation of $S_8$ tension (existed in $\Lambda$CDM)
\cite{Hill:2020osr,Ivanov:2020ril,DAmico:2020ods,Krishnan:2020obg,Vagnozzi:2021gjh,Nunes:2021ipq,Goldstein:2023gnw},
and see
\cite{Poulin:2022sgp,Allali:2021azp,Ye:2021iwa,Alexander:2022own,FrancoAbellan:2021sxk,Clark:2021hlo,Simon:2022ftd,Chacko:2016kgg,Buen-Abad:2022kgf,Reeves:2022aoi,Yao:2023qve}
for the attempts towards the resolutions to $S_8$ tension
(independent of EDE).


The success of EDE is crucially dependent on that it takes effect
only at about matter-radiation equality and then must dilute away
rapidly before recombination to not spoil the fit to CMB. However,
why EDE is exited just before the recombination time is not clear,
called the coincidence problem of EDE \cite{Sakstein:2019fmf}. In
the simplest models of EDE (initially EDE is frozen at
$\phi=\phi_i$ by the Hubble friction), we must have
 \begin{equation}
 |\partial^2_\phi V(\phi_i)|\sim9H_c^2 \sim
 10^{-58}\left(eV\right)^2,
 \label{1}
 \end{equation}
 \begin{equation}
    V(\phi_i)\sim3f_{ede}H_c^2M_p^2\lesssim \left(0.1 eV\right)^4,\label{Vphi}
 \end{equation}
where $M_p=2.435\times10^{27}eV$ is the Planck scale, $H_c$ is the
Hubble parameter at critical redshift $z=z_c$ at the time when the
EDE starts to roll, and $f_{ede}$ is the corresponding energy
fraction of EDE. Thus the initial condition of EDE is very
special. The requirements (\ref{1}) and (\ref{Vphi}) make applying
some well-motivated effective potentials to EDE difficult.

Inflation is the current paradigm of early universe, which
predicts (nearly) scale-invariant scalar perturbation, as well as
primordial gravitational waves. The intensifying of Hubble tension
and the injection of EDE are bringing us different insight into
inflation. It is interesting to note that though based on
$\Lambda$CDM model the Planck collaboration has showed $n_s\approx
0.965$ \cite{Planck:2018jri}, $n_s=1$ ($n_s-1\sim {\cal O}
(0.001)$) is being favored observationally
\cite{Ye:2020btb,Ye:2021nej,Jiang:2022uyg,Smith:2022hwi,Jiang:2022qlj,Peng:2023bik}
in light of the EDE resolution of Hubble tension, see recent
Refs.\cite{Kallosh:2022ggf,Ye:2022efx,Jiang:2023bsz,DAmico:2021fhz,
Takahashi:2021bti,Fu:2023tfo} for its implication for inflation
models. How to unify inflation and EDE in a workable model is an
interesting issue.

It might be thought that EDE is a light field satisfying (\ref{1})
which has stayed $\phi_i$ since the beginning time of inflation
until its unfreezing at $z_c$. According to (\ref{1}),
$|\partial^2_\phi V(\phi_i)|\sim9H_c^2\ll H_{inf}^2$, thus the
fluctuations of EDE during inflation is $\delta\phi\sim
{H_{inf}\over 2\pi}$, where $H_{inf}$ is the inflationary Hubble
parameter, thus we have
\begin{equation}
\left(\delta\phi_i\right)^4\sim H_{inf}^4\gg V(\phi_i)\sim
\left(0.1 eV\right)^4,
\end{equation}
As a result, EDE will be unlikely to have uniform initial values
after inflation. Provided that EDE is initially not at $\phi_i$
but satisfies $V(\phi)>H_{inf}^4\gg V(\phi_i)$, a rolling velocity
of EDE will make it rapidly overshoot $\phi_i$. Thus it is
significant to ask how making EDE settle down at the initial
conditions required by observations.

\begin{figure}[htbp]
    \includegraphics[width=0.9\textwidth]{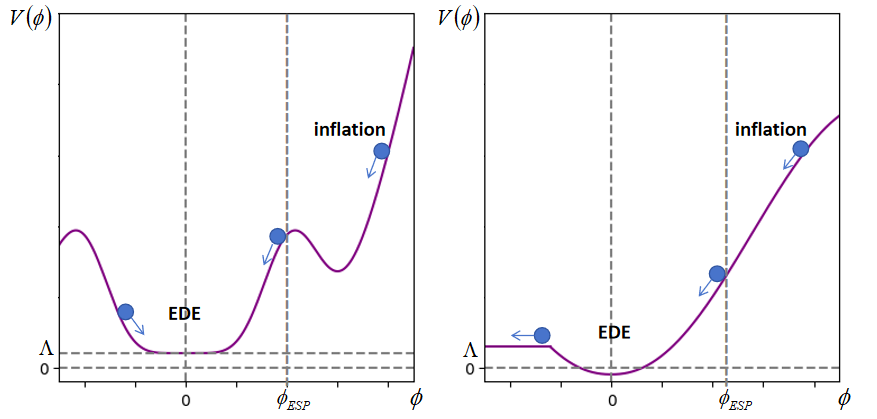}\label{Vi}
\caption{\label{V} The left panel is the sketch of inflation
+axion EDE \cite{Poulin:2018cxd}, in which
$V=V_{inf}+V_{ede}(1-\cos{\phi/f})^3$ with the inflation potential
$V_{inf}$, noting $V_{inf}\gg V_{ede}$. The right panel is that of
inflation +AdS EDE \cite{Ye:2020btb}. It is expected that after
most of inflaton energy decay into radiation, the scalar field
will be trapped near $\phi_{ESP}$ (the vicinity around the initial
value $\phi_i$ of EDE field required by observations) through
particles created, until the field is unfrozen at $z_c$ and EDE
starts to work. After recombination, the universe is assumed to be
$\Lambda$CDM-like, e.g.\cite{Wang:2022jpo}. }
\end{figure}

In this paper, we will present such a mechanism. After inflation,
most of inflaton energy decays into radiation. According to
Ref.\cite{Kofman:2004yc,Green:2009ds}, we consider that the EDE
field passed through \textsf{enhanced symmetry points} (ESP) with
a velocity, see Fig.\ref{V}\footnote{As a model-independent
discussion, we will not distinguish whether EDE is a relic of
inflaton or not, and only focus on the case that after inflation
ended, EDE has a large initial velocity, which is realistic.}.
Particles created at ESP will pull the field back so that $\phi$
can be set near ESP. We will show how this mechanism can work, in
particular how it can efficiently relieve the bound (1), and
discuss the possibility that after inflation ended, EDE can settle
down at the initial conditions required by observations, and the
corresponding particles make a constitution of dark matter (DM).

The role of ESP in the building of inflation model, i.e.trapped
inflation, has been showed in Ref.
\cite{Green:2009ds,Pearce:2016qtn}, see also
e.g.\cite{Piao:2004fh,Brissenden:2023yko} for the current
accelerated expansion. The effects of ESP on the initial
conditions of EDE has been mentioned in
Refs.\cite{Sarkar:2023vpn,Brissenden:2023yko}, but here we
highlighted how the bound (\ref{1}) on EDE can be relieved and the
corresponding particles constitute DM. Their coupling might help
to alleviate $S_8$ tension,
e.g.\cite{Karwal:2021vpk,McDonough:2021pdg,Wang:2022bmk}.

To illustrate this mechanism, consider such a Lagrangian
\cite{Kofman:2004yc,Green:2009ds}
\begin{equation}
\mathcal{L}=\frac{1}{2}\partial_\mu\phi\partial^\mu\phi-V(\phi)+
\frac{1}{2}\partial_\mu\chi\partial^\mu\chi-\frac{1}{2}m_\chi^2\chi^2
-\frac{1}{2}g^2(\phi-\phi_{ESP})^2\chi^2,
\end{equation}
where $m_\chi$ is the mass of $\chi$ particles, which is coupled
to EDE by
$\mathcal{L}_{int}=-\frac{1}{2}g^2(\phi-\phi_{ESP})^2\chi^2$ with
$g$ being the coupling cosntant.

Let $v$ denote the velocity of $\phi$ when $\phi$ passes through
the ESP. In the case of $v/M_p^2\ll g$, $\chi$ particles are
created during a narrow era $|\phi|\lesssim(v/g)^{1/2}$, which
occurs nearly instantaneously. The occupation number $n_k$ of
$\chi$ particles with momenta $k$ is
\begin{equation}
n_k=\exp(-\frac{\pi(k^2+m_\chi^2)}{gv}).
\end{equation}
The corresponding number density of $\chi$ particles is
\begin{equation}
    n_{\chi}|_{T=T_{trap}}=\frac{1}{2\pi^2}\int dkk^2n_k=\frac{(gv)^{3/2}}{(2\pi)^3}\exp(-\frac{\pi m_\chi^2}{gv})
\end{equation}
Supposing $gv\gg m_\chi^2$, we have
\begin{equation}
    n_{\chi}|_{T=T_{trap}}\simeq\frac{(gv)^{3/2}}{(2\pi)^3},\label{nchi}
\end{equation}
where $T_{trap}$ is the temperature of relativistic particle
(radiation) at the time when non-relativistic $\chi$ particles are
produced. Practically, the trapping of EDE may not coincide with
the reheating though it might be part of reheating, so we use
$T_{trap}$ to distinguish it from the reheating temperature $T_r$.




Thus with $\chi$ particles, the original EDE potential is
effectively modified as
\begin{equation}
    V_{eff}(\phi)=V(\phi)+\frac{1}{2}g^2\langle\chi^2\rangle(\phi-\phi_{ESP})^2,
\label{Veff}
\end{equation}
with
$\langle\chi^2\rangle\simeq\frac{1}{2\pi^2}\int\frac{n_kk^2dk}{\sqrt{k^2+m_\chi^2}}\simeq\frac{n_\chi}{m_\chi}$.
Thus it is possible that we have $\partial_\phi V_{eff}=0$ and
$\partial^2_\phi V_{eff}(\phi)\gtrsim0$, equivalently
\begin{equation}
\partial_\phi^2V(\phi_T)+g^2\langle\chi^2\rangle(\frac{T}{T_{trap}})^3\gtrsim0.
\label{trap}
\end{equation} at certain $\phi_T$, noting
$\langle\chi^2\rangle_T=\langle\chi^2\rangle(\frac{T}{T_{trap}})^3$
with $n_\chi\propto T^3$. Thus after $\chi$ particles are created,
the EDE field that has passed through ESP will be pull back and
naturally trapped at $\phi_T$. As the universe expands,
$T$ lowered, the corresponding trapping effect will be inevitably
weaken.


The condition that at matter-radiation equality EDE started
rolling is
\begin{equation}
|\partial^2_\phi V(\phi_i)|\sim
9H_c^2+g^2\langle\chi^2\rangle(\frac{T_{ede}}{T_{trap}})^3
\label{init}
\end{equation}
if $\partial^2_\phi V(\phi_i)<0$. It is possible that the Hubble friction dominated at $z_c$,
$g^2\langle\chi^2\rangle(\frac{T_{ede}}{T_{trap}})^3\ll9H_c^2$,
since the trapping effect is weaken as time flows. In this case,
the physics of EDE is unchanged and similar to those in
Refs.\cite{Poulin:2018cxd,Agrawal:2019lmo,Ye:2020btb}, and we will
have the unfrozen condition (\ref{1}).


However, if
$g^2\langle\chi^2\rangle(\frac{T_{ede}}{T_{trap}})^3\gtrsim9H_c^2$,
the trapping effect will be significant at $z_c$. It can be
expected that EDE will be set at $\phi_i$ at $T_{ede}$ if
(\ref{trap}) is satisfied, however, the energy density of EDE at
$\phi_i$ is slightly lower than that at $\phi_T$ ($T>T_{ede}$),
due to the slope of $V(\phi)$ in (\ref{Veff}). Combining
$g^2\langle\chi^2\rangle(\frac{T_{ede}}{T_{trap}})^3\gtrsim9H_c^2$
and (\ref{Vphi}), we have
\begin{equation}
g^2\langle\chi^2\rangle(\frac{T_{ede}}{T_{trap}})^3\gtrsim{3V(\phi_i)\over
f_{ede}M_p^2}.\label{ft}
\end{equation}
where $T_{ede}\sim10^{-11}GeV$ is the energy scale of EDE. It is
interesting that $\chi$ might be a part of DM, thus
\begin{equation}
    n_\chi m_\chi(\frac{T_0}{T_{trap}})^3\sim\lambda\Lambda,\label{dm}
\end{equation}
where $T_0\sim10^{-13}GeV$ is the radiation temperature at present
and $\Lambda\sim(10^{-4}eV)^4$ is the cosmological constant, and
$0<\lambda<1$ corresponds to the energy fraction of $\chi$
particles in DM. Combining (\ref{nchi}) and (\ref{dm}), we have
\begin{equation}
m_{\chi}\sim\lambda\frac{(2\pi)^3}{(gv)^{3/2}}T_{trap}^3(10^{-4}eV).\label{mchi}
\end{equation}
According to (\ref{trap}) and (\ref{dm}), we have
\begin{equation}
g\gtrsim44\lambda^{1/5}\left(\frac{T_{trap}}{v}10^{-20}eV\right)^{3/5},
\label{trapdm}
\end{equation}
which is the requirement for EDE trapped. According to (\ref{ft})
and (\ref{dm}), we have
\begin{equation}
     g\gtrsim44\lambda^{1/5}\left(\frac{T_{trap}^2}{v}10^{-18}\right)^{3/5}\label{ftdm}
\end{equation}
which is the requirement for EDE rolling at $z_c$. It is noted
that (\ref{ftdm}) is actually tighter than (\ref{trapdm}).
According to (\ref{ftdm}), for an instant $T_{trap}\sim
T_r\sim10^{15}GeV$, as well as $\lambda=1$ ($\chi$ particles
contribute all DM) and $v\sim10^{-8}M_p^2$ \footnote{It is obvious
that our results depend on the assumptions for the values of $v$
and $T_{trap}$. It is assumed that the reheating happened before
the trapping of EDE, thus the constraint on $T_{trap}$ and $v$ are
$T_{trap}\lesssim T_r\lesssim 10^{15}$Gev and $v\lesssim
10^{-8}M_p^2$. }, we have
\begin{equation}
g\gtrsim10^{-9}.
\end{equation}

The slope of EDE potential is related to the coupling $g$ of EDE
with $\chi$ particles by
\begin{equation}
\partial^2_\phi V(\phi_i)\sim
\frac{1}{(2\pi)^6}{g^5v^3\over \lambda T_{trap}^{6}}
(0.1eV)^2.\label{init2}
\end{equation}
Thus the bound (\ref{1}) can be effectively relaxed for a stronger
coupling $g$, a larger velocity at ESP, or a lower $T_{trap}$.
Generally, $g\ll 1$ is required, thus that trapped EDE is workable
suggests
\begin{equation}
v\gtrsim10^{-21}M_p^2. \end{equation}

\begin{figure}[htbp]
    \includegraphics[width=0.9\textwidth]{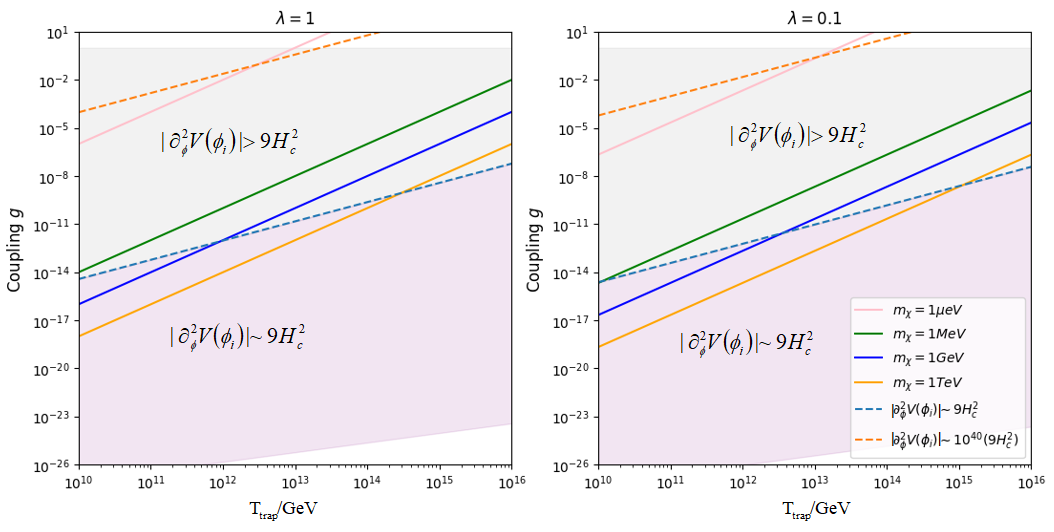}
\caption{\label{Tr}The compatible parameter space for the coupling
$g$ and the temperature $T_{trap}$ of radiation at trapping time,
in which $m_\chi=$1MeV (green solid), 1GeV (blue solid) and 1TeV
(orange solid). The trapping effect dominated at $z_c$, i.e.
$|\partial^2_\phi V(\phi_i)|>9H_c^2$, which corresponds to the
grey region, and the Hubble friction dominated at $z_c$, i.e.
$|\partial^2_\phi V(\phi_i)|\sim9H_c^2$, which corresponds to the
purple region, both are divided by the blue dashed line, changes
as $T_{trap}$ varies. The orange dashed line is that with
$|\partial^2_\phi V(\phi_i)|\sim 10^{40}H_c^2$ for $g\sim0.1$
around $T_{trap}\sim10^{13}GeV$. }
\end{figure}

In Fig.\ref{Tr}, we show the allowed $g$ with respect to
$T_{trap}$ for certain $m_\chi$ assuming $v\sim10^{-8}M_p^2$,
$\lambda=1$ and $\lambda=0.1$ respectively. The bound (\ref{1})
can be relieved only if
$g^2\langle\chi^2\rangle(\frac{T_{ede}}{T_{trap}})^3\gtrsim9H_c^2$.
In this case, the mass range of $m_{\chi}$ beyond $1$TeV is ruled
out, since it needs $T_{trap}\gtrsim 10^{15}$GeV.

It is also significant to note that the lower $T_{trap}$ will
allow the EDE potential steeper, $|\partial^2_\phi V(\phi_i)|\gg
9H_c^2$. In particular, for the strongest coupling $g\sim0.1$
around $T_{trap}\sim10^{13}GeV$, $|\partial^2_\phi V(\phi_i)|\sim
10^{40}H_c^2$ (while $\chi$ particles have mass
$m_\chi\sim10^{-6}$eV). The relief for (\ref{1}) is effective as
long as the mass of $\chi$ particles is small enough, whether allowed
$\chi$ to contribute all DM or not.

\begin{figure}[htbp]
    \includegraphics[width=0.5\textwidth]{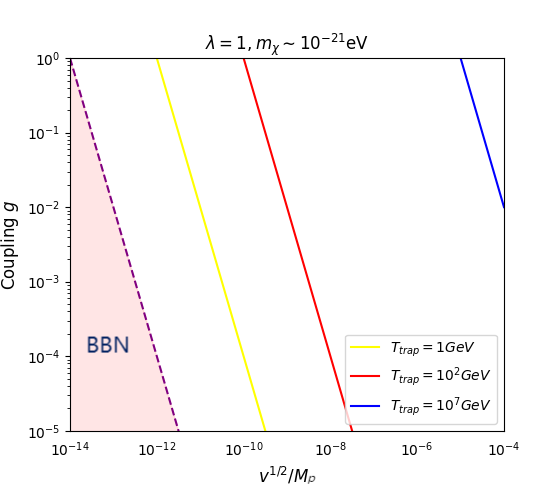}
\caption{\label{m} The compatible parameter space for the coupling
$g$ and the velocity $v$ of EDE at ESP, with $\lambda=1$ ($\chi$
particles constitute all DM) and $m_\chi\sim10^{-21}$eV. The
models represented by yellow, red and blue solid lines correspond
to $T_{trap}=1$GeV, 100GeV, and $10^7$GeV, respectively. The
models in the red region is forbidden because of $T_{trap}$ lower
than the BBN temperature. }
\end{figure}

In Fig.\ref{m}, we show the conditions that $\chi$ particles can
have very small mass, in particular $m_\chi\sim10^{-21}$eV. Recent
observations suggest that the scalar field DM should have the mass
$m_\chi>2.3\times10^{-21}$eV,
e.g.\cite{Irsic:2017yje,Armengaud:2017nkf,Nori:2018pka}. As we see
in Fig.\ref{m}, $T_{trap}\lesssim 10^7GeV$ is necessary for
$g\lesssim 0.1$, or it required $v\gtrsim 10^8M_p^2$.


Indeed, there are lots of mechanisms to create DM when inflation
ended
\cite{Chung:1998zb,Chung:2001cb,Markkanen:2018gcw,Fairbairn:2018bsw,Chung:2018ayg,Hashiba:2018tbu,Herring:2019hbe,Kainulainen:2022lzp,Garcia:2022vwm},
The cases with a low reheating temperature also has been
intensively studied
\cite{Giudice:2000ex,Fornengo:2002db,Pallis:2004yy,Silva-Malpartida:2023yks},
usually triggered by the decay of a long-lived massive particle.
Trapped EDE could also accommodate with them if the $\chi$
particles contribute to a part of DM.


In summary, we investigated a DM-trapped EDE mechanism, by which
not only after inflation ended EDE can naturally settle down at
the initial conditions required by recent observations, but also
the bound (\ref{1}) on the EDE potential is relieved, in
particular EDE might be workable for a steeper potential,
$|\partial_\phi^2V(\phi_i)|^{1/2}\gg H_c$. As displayed in
Fig.\ref{Tr}, it is compatible to relieve the bound (\ref{1}) in a
broad range of relevant parameters. The relief of the bound
(\ref{1}) is beneficial to enrich the building of EDE models.
Theoretically the preference at ESPs might suggest a spontaneous
origin for the initial conditions of EDE.

There are much open issues left. The MCMC analysis for DM-trapped
EDE, especially with a possible steeper EDE potential than
(\ref{1}), needs to be performed, which is currently under
investigation. The coupling EDE with DM might have a unforeseen
role in mitigating $S_8$ tension,
e.g.\cite{Karwal:2021vpk,McDonough:2021pdg,Wang:2022bmk}. It is
also interesting to consider to trap EDE with relativistic
particles. In EDE cosmologies the spectral tilt $n_s$ of scalar
perturbation will shift towards $n_s=1$, how to unify the
corresponding inflation and EDE in a well-motivated model is still
a challenge. It might be also possible that during inflation EDE
corresponds to one lower energy vacuum in multiple-vacua
landscape, which is associated with birth of supermassive
primordial black holes, e.g.\cite{Huang:2023chx,Huang:2023mwy}.
Recently, it has been showed that EDE could be implemented
naturally in string theory
\cite{McDonough:2022pku,Cicoli:2023qri,Cicoli:2023opf}, thus the
moduli trapping in corresponding contexts is also worth studying.

\section*{Acknowledgments}

We thank Jun-Qian Jiang for discussions. This work is supported by
NSFC, No.12075246, National Key Research and Development Program
of China, No. 2021YFC2203004, and the Fundamental Research Funds
for the Central Universities.

\end{document}